# Media Moments and Corporate Connections: A Deep Learning Approach to Stock Movement Classification

**Matthew Sahagun***
mevsahagun@gmail.com

**Luke Sanborn***
luke@steepbrook.com

## Abstract

The financial industry poses great challenges with risk modeling and profit generation. These entities are intricately tied to the sophisticated prediction of stock movements. A stock forecaster must untangle the randomness and ever-changing behaviors of the stock market. Stock movements are influenced by a myriad of factors, including company history, performance, and economic-industry connections. However, there are other factors that aren't traditionally included, such as social media and correlations between stocks. Social platforms such as Reddit, Facebook, and X (Twitter) create opportunities for niche communities to share their sentiment on financial assets. By aggregating these opinions from social media in various mediums such as posts, interviews, and news updates, we propose a more holistic approach to include these "media moments" within stock market movement prediction. We introduce a method that combines financial data, social media, and correlated stock relationships via a graph neural network in a hierarchical temporal fashion. Through numerous trials on current S&P 500 index data, with results showing an improvement in cumulative returns by 28%, we provide empirical evidence of our tool's applicability for use in investment decisions.

## 1 Introduction

Stock prices have an inherent nature of volatility, creating difficulty in predicting their peaks and troughs. Investing in markets with an unknown, non-stationary trait results in great risk when attempting to collect profits. Numerous factors alter the prices, such as company history/performance, and economic-industry connections. Although common amongst most informed investors, naivety increases financial risk, resulting in losses within the stock market. As fundamental and technical analyses have been founded to attempt to encapsulate the market, they fail to incorporate unexpected/unforeseen "media moments" around the world. Media moments - tweets, Reddits, news articles, Facebook, or anywhere a post has occurred - have a monumental influence on the stock market. For example, CEO Elon Musk expressed beliefs about Tesla's high stock price, claiming the stock was overvalued. The result was a $13 Billion loss to the company's market capitalization.

Financial markets are informationally efficient: all information pertaining to a company's stock is incorporated into its current price. (Malkiel, 1989). Current implementations of data focus on stock-relevant data, neglecting the opportunity to maximize learning holistically through media moments and interrelated stocks, resulting in an incomplete understanding of how stock movement occurs. Multimodal stock analysis faces numerous obstacles. For example, price signals and media moments are sometimes disjointed or in conjunction. A price signal paired with a media moment can either create a unified context or opposing perspectives. This is due to the differences in price signals and media moment influence. Many media moments cause disruption to a company, while price signals may not produce as much noise, and vice versa. Continuing on the efficient market hypothesis, we present **ATOMIC-SM: Attention-Driven Temporal Order Modeling for Interdependent Companies via Stock Movement Classification**. ATOMIC-SM trains a Gra;ph Attention Network for stock prediction based on hierarchical attention capture. ATOMIC-SM simultaneously learns from media moments and price signals (along with fundamental and technical analysis) for stock prediction. Through multiple trials, we show ATOMIC-SM's ability to forecast, leading to probability analysis within high-risk scenarios.

---

* Equal contribution

## 2   Problem Formulation

ATOMIC-SM's prioritized objective is to understand temporally relevant information (within media moments) by comprehending the online environment to dissipate biased/niche events, creating accurate predictions in stock movements.

## 3   ATOMIC-SM Components and Learning

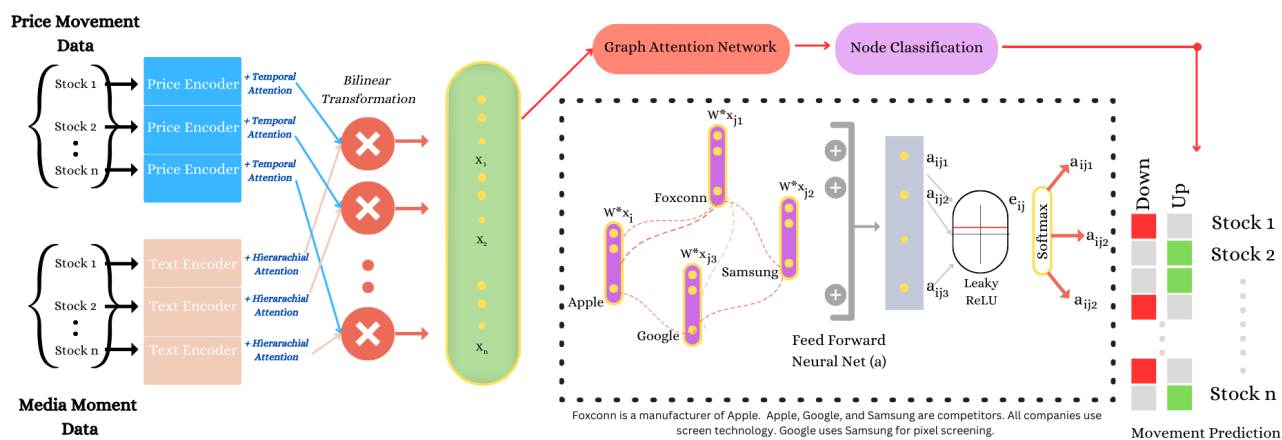

Figure 1: An overview of ATOMIC-SM: Encoding Mechanisms, GAT Mechanism, Joint Optimization.

In this section, we will give an overview of ATOMIC-SM, followed by a detailed description of each component. ATOMIC-SM encodes stock features for each trading day over a fixed period of time.

### 3.1   Technical Encoder

In our research paper, we introduce the Technical Encoder component, which is a crucial part of our ATOMIC-SM model. The Technical Encoder is responsible for encoding historical price movements of stocks, providing valuable insights into their temporal trends. We use the adjusted closing price, highest price, and lowest price for each trading day as input features. These features are then normalized with the previous day to ensure that the model captures price movements rather than absolute price values.

To model the sequential dependencies across trading days, we employ a Long Short-Term Memory (LSTM) network. The LSTM output for each day is represented as $h_i = LSTM(p_i, h_{i-1})$, where $h_i$ denotes the hidden states of the LSTM. These hidden states capture the temporal trends in stock prices, which are essential for predicting future movements.

Furthermore, we introduce a temporal attention mechanism to weigh the importance of each trading day's information in capturing stock trends. This mechanism, denoted as ζ(·), learns to assign adaptive weights to each day's hidden state, rewarding days with more impactful information. The temporal attention mechanism is defined as follows:

$$\beta_i = \frac{exp(h_i^T \cdot W_h)}{\sum_{i=1}^{T} exp(h_i^T \cdot W_h)}$$

Here, βi represents the learned attention weights for the trading day $i$, and $W_h$ is a learnable parameter matrix. The temporal attention mechanism aggregates impactful information across all days in the lookback window to produce price features $q_t \in R_v$

## 3.2 Media Moments (Social Media Encoder)

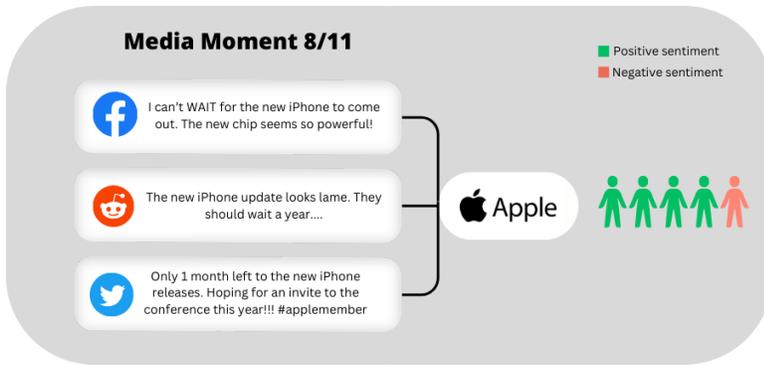

Figure 2: An overview of Media Moments

In our ATOMIC-SM model, we aim to capture the influence of social media on stock movements by introducing the Media Moments Encoder. This component is responsible for extracting features from social media content, such as tweets from Twitter, posts from Reddit, and discussions on platforms like LinkedIn and Facebook. Social networks have been shown to convey user sentiment toward financial commodities beyond price trends (Xu and Cohen 2018).

We calculated a sentiment score for each company, ranging from 0 to 1, representing the positive sentiment associated with that company. This sentiment score is derived from social media content, including tweets, posts, and discussions across various platforms. The activity score measures the social media activity related to each company. It is calculated by a combination of mentions and discussions aggregated from various social media platforms, including Reddit, Twitter, LinkedIn, and other relevant sources. The goal is to quantify the level of attention and engagement a company receives on social media.

To ensure consistency and comparability of these scores across different companies and timeframes, we apply a normalization process with the previous day's scores as follows: $score = \frac{S_i}{S_{i-1}}$.

## 3.3 Blending Multimodal Information

In our ATOMIC-SM model, we recognize the complementary nature of information from different modalities, such as historical prices and media moments from social networks (Robert P. Schumaker, 2019). To effectively blend this multimodal information, we use a bilinear transformation that captures the pairwise feature interactions between price features ($q_t$) and media moment features ($c_t$). (Sawhney and Agarwal, 2020)

## 3.3 Graph Attention Network

In our research paper, we recognize that stocks are interconnected and their relationships can significantly impact each other. To model these interconnections, we represent stocks and their relations as a graph and introduce a Graph Attention Network (GAT) component. We leverage Wikidata company profiles to create meaningful relationships between stock nodes. Other approaches (Feng et al. 2019) use properties that might create false, irrelevant ties between stocks and distract from the critical measures.

| Wikidata ID | Relationship | Description |
| --- | --- | --- |
| P127 | "owned by | Indicates ownership relationships between companies in the stock market. |
| P355 | "subsidiary of" | Represents the hierarchical structure of ownership where one company is a subsidiary of another. |
| P836 | "controlled by | Denotes control relationships between companies, which can influence stock dynamics. |
| P1553 | "has part" | Specifies that one company has a part that is relevant to another. |
| PContract ID | "has contract with" | Represents contractual relationships between companies that can influence their stock movements. |

Table 1: Top 6 most impactful relationships between companies ordered greatest to least.

We create the stock relation network as a graph G(S, E), where S represents the set of nodes (stocks), and E represents the set of edges (relations between stocks). We extract first and second-order relations between stocks using Wikidata, allowing us to establish connections based on company relationships.

The GAT layer in our model takes the encoded multi-modal market information as input and produces updated node features using a self-attention mechanism. The self-attention mechanism assigns attention coefficients αij to each pair of stocks i and j, representing the importance of their relations. These attention coefficients are computed as follows:

$$\alpha_{ij} = exp(LeakyReLU(a^T \cdot [Wx_i \oplus Wx_j])$$

$$\sum_{k \in N_i} exp(LeakyReLU(a^T \cdot [Wx_i \oplus Wx_j])$$

Here, α is a learnable weight matrix and $N_i$ represents the immediate neighborhood of node $i$. These attention coefficients guide the aggregation of features from neighboring stocks, allowing the model to capture the relevance of relations in predicting stock movements.

Our ATOMIC-SM model incorporates these components to effectively blend multimodal information from historical prices, social sentiment, and inter-stock relations, providing a comprehensive approach to stock movement prediction.

## 4  Experiments

### 4.1  Data collection

We collected data for stocks with a market capitalization between two billion and one trillion dollars within the Nasdaq. ATOMIC-SM was trained on data from January 2019 to March 2021 with approximately 600 trading days containing layered data analysis. Data was divided 70/15/15 between training, validation, and testing subsets respectively. Sequences of 5 trading days were classified either as a positive price movement or a negative price movement, creating 261,390 individual data points. The dataset has a ratio of 48.21:51.87 between the two classes respectively. Company relationships were updated per year of training data, reflecting changes between January 2019, January 2020, and January 2021 for those stocks. We used SocialSentiment.io for the social media score and activity for each respective stock in the dataset.

Since our data was collected between medium-market capitalization and large-market capitalization stocks, any penny stock was disqualified from being evaluated. Penny stocks are known for their incredulous volatility, creating false hopes for inexperienced investors. In addition, penny stocks do not represent the stock market well, as they are solely driven by media moments. Therefore, mid-cap and large-cap stocks represented comprehensive volatility.

### 4.2  Model Setup

During the training phase of our ATOMIC-SM model, several crucial parameters were carefully configured to optimize its performance. The model employed Graph Attention Networks (GAT) with a total of 4 attention heads, allowing it to effectively capture complex relationships among stocks in the market. To facilitate effective learning, the model utilized a learning rate of 4e-4, which determined the step size during the optimization process. For the temporal aspect of our model, we used LSTM units with a hidden size of 64.

The training process spanned 8,000 epochs, providing ample opportunities for the model to learn and adapt to the data. However, to prevent overfitting and ensure that the model generalizes well to new data, we implemented early stopping. Early stopping was based on the F1 score on the validation dataset, allowing us to halt training when the model's performance ceased to improve significantly. To efficiently process and update the model's weights, a batch size of 8 was utilized. Lastly, the sequence length of the data used for training was set to 5 trading days. This sequence length allowed the model to consider patterns and relationships over a specific time frame, aligning with the nature of stock market data where short-term trends are essential for prediction. We adopted a trading strategy where we would buy the top 4 most confident positive price movement predictions. Those stocks were then sold at the market opening the next day.

## 4.3 Evaluation

In our evaluation of the ATOMIC-SM model, we used the Matthews Correlation Coefficient (MCC) in addition to the F1 score. The MCC is particularly valuable when dealing with imbalanced datasets, as it considers true positives, true negatives, false positives, and false negatives, providing a more comprehensive assessment of the model's performance. MCC accounts for the balance between sensitivity (true positive rate) and specificity (true negative rate), making it a robust measure for binary classification tasks like stock movement prediction. A higher MCC indicates better predictive performance, with values ranging from -1 (perfect inverse prediction) to +1 (perfect prediction) and 0 indicating no better than random prediction.

$$MCC = \frac{TP*TN - FP*FN}{\sqrt{(TP+FP)(TP+FN)(TN+FP)(TN+FN)}}$$

When transitioning from model evaluation to real-world testing, it's crucial to recognize that performance is highly dependent on market conditions. The financial market is inherently volatile and influenced by various macroeconomic factors, geopolitical events, and market sentiment. As a result, model performance may vary significantly based on the prevailing market condition. For our real-world test set, we took the same 87 stocks the model was trained on and fed it data from March to December 2021. This distinct time period avoids any leaks from training/validation/testing data into the real-world dataset and makes sure that this is novel data for ATOMIC-SM.

In our real-world market test, we used the Sharpe coefficient as one of the key evaluation metrics. The Sharpe coefficient is a measure of the risk-adjusted return of an investment or trading strategy. It considers both the returns generated and the level of risk associated with those returns. The Sharpe coefficient helps translate ATOMIC-SM's potential into financial markets by accounting for risk in addition to profitably. It is arguably more important to have a less volatile strategy with more consistent returns than a very risky portfolio.

To provide a benchmark for our model's performance, we compared its results to those of the S&P 500 ETF known as SPY. The SPY ETF closely tracks the performance of the S&P 500 index, making it a widely recognized benchmark for assessing the overall performance of the U.S. stock market. By comparing our model's performance to the SPY ETF, we aimed to gauge its ability to generate returns that outperform or align with a standard market index, offering investors a valuable reference point for decision-making.

## 4.4 Baseline Comparison

We compared ATOMIC-SM with the following baselines:
- Random Guess: Random guess as price rises or falls
- ARIMA: Autoregressive Integrated Moving. Average models historical prices as a nonstationary time series (Brown, 2004).
- RandForest: Random Forests classifier trained over word2vec embeddings for tweets.

- TSLDA: Topic Sentiment Latent Dirichlet Allocation model is a generative model that uses sentiments and topic modeling on social media (Nguyen and Shirai, 2015).
- HAN: A hierarchical attention mechanism to encode textual information during a day and across multiple days (Hu et al., 2018).
- StockNet: A variational Autoencoder (VAE) that uses price and text information. Text is encoded using hierarchical attention during and across days. Price features are modeled sequentially (Xu and Cohen, 2018).
- HATS - A hierarchical graph attention network for stock movement prediction. (Kim et al., 2019)
- Adversarial LSTM - Machine learning model for stock movement prediction trained via intentional perturbations to simulate the stochasticity of price variables. (Feng et al., 2019a)
- MAN-SF - A hierarchical graph attention network trained via fundamental & technical analysis. Trained on Twitter posts to show sentiment within online content's influence on stock movement. (Sawhney and Agarwal, 2020)

| Model | F1 | Accuracy | MCC |
|---|---|---|---|
| Random Guess | 0.402 ± 7e−4 | 0.394 ± 7e−4 | 0.013 ± 1e−3 |
| ARIMA (Brown, 2004) (Technical analysis only) | 0.513 ± 1e−3 | 0.514 ± 1e−3 | −0.021 ± 2e−3 |
| RandForest (Venkata Sansake Pagolu, 2016) (Fundamental analysis only) | 0.527 ± 2e−3 | 0.531 ± 2e−3 | 0.013 ± 4e−3 |
| TSLDA (Nguyen and Shirai, 2015) | 0.539 ± 6e−3 | 0.541 ± 6e−3 | 0.065 ± 7e−3 |
| HAN (Huet al., 2018) (Fundamental analysis only) | 0.572 ± 4e−3 | 0.576 ± 4e−3 | 0.052 ± 5e−3 |
| StockNet - HedgeFundAnalyst (Xu and Cohen, 2018) | 0.575 ± − | 0.582 ± − | 0.081 ± − |
| HATS (Kim et al., 2019) (Fundamental analysis only) | 0.560 ± 2e−3 | 0.582 ± − | 0.081 ± − |
| Adversarial LSTM (Feng et al., 2019a) | 0.570 ± − | 0.572 ± − | 0.148 ± − |
| MAN-SF (Sawhney and Agarwal, 2020) | 0.605 ± 2e−4 | 0.608 ± 2e−4 | 0.195 ± 6e−4 |
| **ATOMIC-SM (This work)** | **0.663 ± 3e-4** | **0.682 ± 3e-4** | **0.198 ± 9e-4** |

Table 2: Results Compared with Baselines. Bold shows the best results. Numbers closer to 1 represent higher performance.

## 5   Results and Analysis

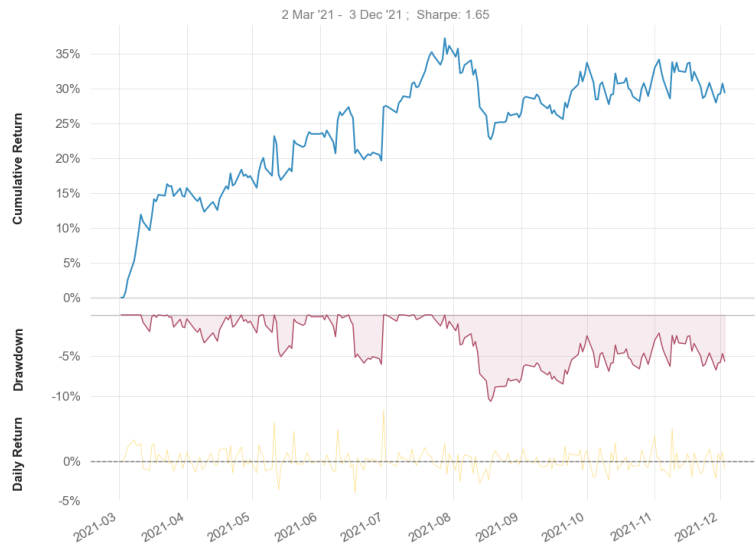

Figure 3: ATOMIC-SM performance from March 2021 to December 2021

|  | Cumulative Returns | Sharpe Ratio |
|---|---|---|
| Benchmark: SPY Index | 18.07% | 1.88 |
| ATOMIC-SM | 29.47% | 1.65 |

Table 3: ATOMIC-SM and benchmark performance from March 2021 to December 2021

### 5.1   Analysis of Media Moments

| Discussed Stock on r/WallStreetBets | Symbol | Days the Stock Was In The Top 5 Most Mentioned |
|---|---|---|
| GameStop | SME | 261 |
| Tesla | TSLA | 202 |
| AMC Entertainment | AMC | 151 |
| Palantir Technologies | PLTR | 135 |
| BlackBerry Limited | BB | 68 |

Table 4: Most Discussed stocks on r/WallStreetBets[1]

The data depicted in the chart highlights a prevalent phenomenon in niche communities like r/WallStreetBets, where certain stocks are consistently discussed and overrepresented in discussions. This overrepresentation can lead to skewed investor sentiment, potentially distorting the overall view of stock market dynamics. To address this challenge and gain a more comprehensive understanding of investor sentiment, it is imperative to explore a wider spectrum of social media platforms. These platforms often attract different demographic groups of users with diverse needs and discussions.

---
[1] https://capital.com/wallstreetbets-returns

By extending our analysis beyond a single subreddit or platform in general, we can tap into the collective wisdom and insights of a more diverse and varied user base. Different platforms offer unique perspectives. Users may engage in discussions that encompass a broader range of stocks and financial topics. This approach not only helps in reducing biases but also provides a more holistic view of investor sentiment and market trends. In table 2, we observed that including these media moments boasted better performance than the baselines. While the MCC did not demonstrate a

Platforms like Twitter, LinkedIn, and other specialized forums cater to distinct audiences with varying interests. Exploring these platforms allows us to capture a more nuanced and balanced sentiment analysis, helping us overcome the limitations of overrepresentation in niche communities.

## 5.2 Conclusion

We propose ATOMIC-SM, a model for stock movement prediction that builds upon prior work and focuses on a holistic overview of media moments. As Table 3 shows, ATOMIC-SM achieved a greater performance than prior MLMs (Machine Learning Models) on the baselines of F1, MCC, and accuracy metrics. We noticed that including a broad range of media moments across several social media platforms increases the models' performance on the metrics aforementioned. By adopting a holistic approach, and aggregating data from various social media platforms and niche communities, we aim to provide a more comprehensive and balanced perspective on stock-related discussions. Furthermore, filtering the relations between companies has allowed for higher attention scores between similar stocks and more diverse hidden correlations. In our 9-month real-world test, ATOMIC-SM outperformed the market benchmark while having a slightly reduced Sharpe coefficient. While an exact quantitative and qualitative representation of the stock market is (and will be) impossible to achieve, market sentiment via social media provides a feasible way to represent such opinions with far fewer orders of magnitude. We have observed that creating datasets with all available information becomes difficult as more online content shifts away from written work. We plan to look at verbal communication on other social media platforms such as YouTube to increase the depth of sentiment analysis.